\pdfoutput=1
\documentclass[11pt]{article}

\usepackage[T1]{fontenc}
\usepackage{lmodern}
\usepackage[letterpaper,margin=1in]{geometry}
\usepackage{amsmath,amssymb,amsfonts}
\usepackage{graphicx}
\usepackage{booktabs}
\usepackage{array}
\usepackage[table]{xcolor}
\usepackage{siunitx}
\usepackage{cite}
\usepackage{microtype}
\usepackage{flafter}
\usepackage[section]{placeins}
\usepackage[hidelinks]{hyperref}

\hypersetup{
  pdftitle={Current-Limiting Control for Fault Ride-Through of LLC-based Solid-State Transformer in Data Centers},
  pdfauthor={Haoyu Wang, Chi Zhang, Mafu Zhang, Rudy Wang, and Peter Barbosa},
  pdfkeywords={current limiting, data center, DC-DC control, LLC converter, short-circuit fault, solid-state transformer}
}

\title{Current-Limiting Control for Fault Ride-Through of LLC-based Solid-State Transformer in Data Centers}
\author{Haoyu Wang \quad Chi Zhang \quad Mafu Zhang\\
Rudy Wang \quad Peter Barbosa%
\thanks{This work has been submitted to the IEEE for possible publication.
Copyright may be transferred without notice, after which this version may no longer be accessible.}}
\date{}

\begin{document}
\maketitle

\begin{abstract}
Solid-State Transformers (SSTs) are increasingly proposed as the interface between distribution grids and data centers due to flexible power flows and fast dynamic response. However, when a short-circuit fault occurs in a load branch, the SST with a voltage-source-type DC-DC stage is forced to shut down due to fault currents. Therefore, current-limiting strategies are strongly needed to prevent catastrophic equipment damage and cascading blackouts by instantly restricting massive current spikes and offering sufficient currents for protection devices to act at the faulted branch. This paper proposes a coordinated DC load fault-tolerant current-limiting and recovery strategy embedded directly in the control of the SST DC-DC stage, avoiding additional hardware cost. Specifically, the fault mechanism of an example LLC resonant converter is studied. Accordingly, a fault detection framework is implemented, a closed-loop current controller is proposed to limit the DC current to a designated value within microseconds by surging the switching frequency and adjusting the duty cycle, and a ramped recovery stage will then restore the DC bus after the fault isolation without inrush currents. Experiments on an LLC converter prototype have verified the feasibility of the proposed current-limiting strategy, enabling faster and lower-cost fault response suitable for resilient data center power architectures.
\end{abstract}


\section{Introduction}

\begin{figure}[t]
\centering
\includegraphics[width=0.74\textwidth]{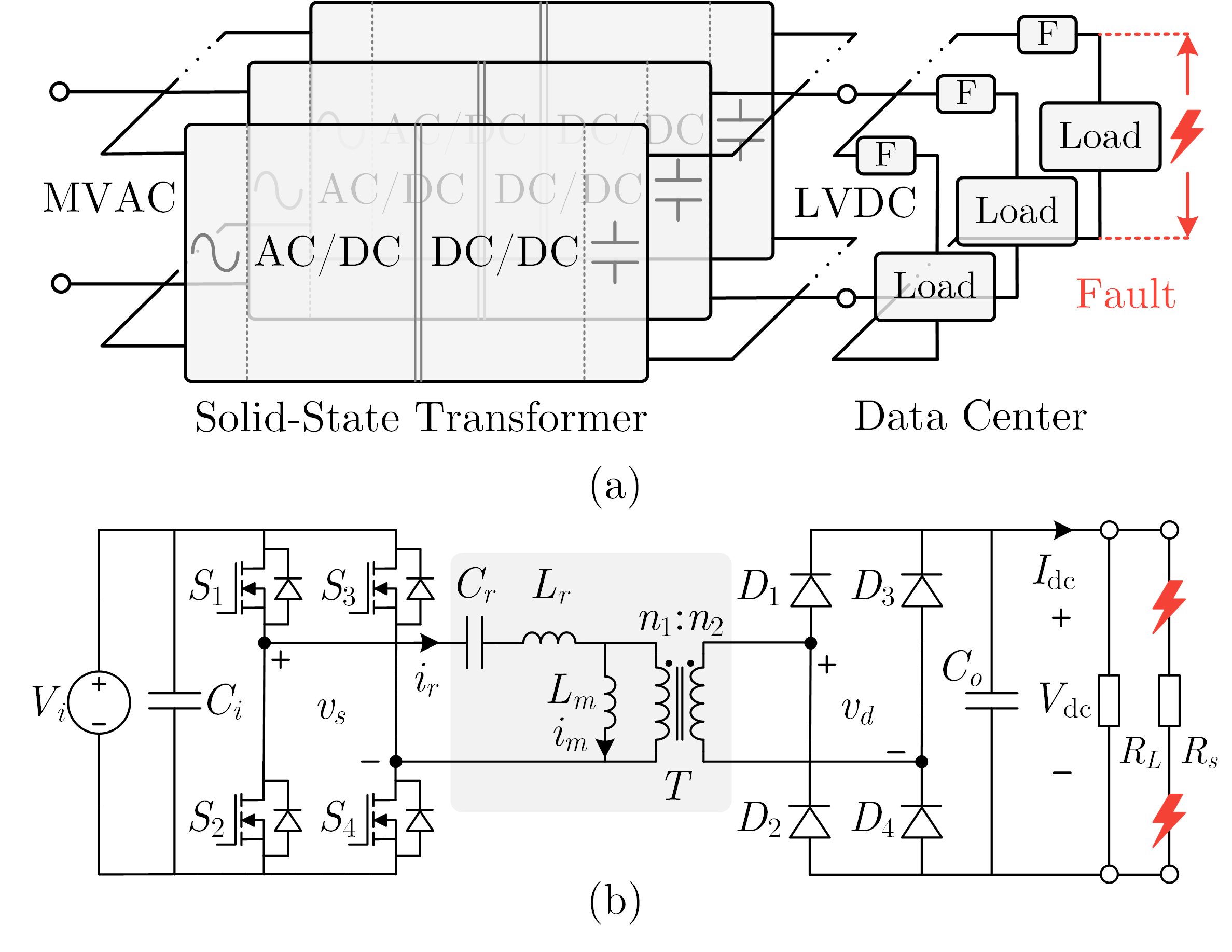}
\caption{System diagram with DC load faults. (a) SST in Data Centers. $\mathrm{F}$ denotes protection devices like fuses. (b) An example of voltage-source-type LLC converters as the DC-DC stage.}
\label{fig:system}
\end{figure}

The rapid growth of artificial intelligence and cloud computing has driven the power demand of data centers to unprecedented levels, motivating a shift toward medium-voltage AC (MVAC) to low-voltage DC (LVDC) architectures, with industry roadmaps targeting in-rack distribution. Solid-state transformers (SSTs) with higher efficiency, reduced conversion stages, and improved compatibility, are well positioned to serve as the grid interface for such DC buses \cite{5634051, 9893537, 10093060, 10242261, 10132875, 10194423, 10319663, 10362302, 9628583, 11358392, 11077776, 10583830}. In many of the SST architectures, the isolated DC-DC stage is realized with voltage-source-type output behavior, such as LLC resonant converters \cite{6648465}, as shown in Fig.~\ref{fig:system}.

A key challenge is that a short-circuit fault occurring at a downstream load (e.g., rack level) collapses the load impedance and drives fault currents to rise within tens of microseconds\cite{app12010015}. For a voltage-source-type DC-DC stage like LLC converters attempting to sustain the bus voltage across the collapsed impedance, an unmitigated fault can drive the resonant current well beyond its rated value before any breaker-based protection, forcing the system to shut down and causing fatal damage to data centers \cite{4118335}. Load branches are typically protected by fuses or circuit breakers that are cleared by $I^2t$ characteristics, selective isolation thus requires the current to be neither unlimited nor blocked, but regulated at a designated level for a specified period of time \cite{9526632}. Therefore, there is growing interest in current-limiting strategies that can provide substantial currents for protection devices to ride through the fault without system damage.

Existing approaches to load-side fault management generally fall into three categories: (1) hardware-based protection in series with the feeder (e.g., DC circuit breakers and fault current limiters) that adds cost \cite{1331495, app12010015}; (2) converter-level over-current protection that clamps the tank to optionally reconstruct the resonant current from tank voltages, which does not hold a specified current or address restoration
\cite{1179276, 4118335}; (3) Control-level fault ride-through that temporarily reconfigures control architecture and keeps the SST a controllable source for the MV side during a fault \cite{9343854, 7482715} or without a closed form \cite{6953876}. The third category without additional hardware is attractive for data centers, allowing the SST to isolate only the faulted load branch, but there lacks a comprehensive current-limiting strategy for SSTs.

Building on the control-level philosophy, this paper proposes a coordinated current-limiting and recovery strategy embedded at the control level of the DC-DC stage of data-center-facing SSTs, especially for the voltage-source types, targeting a downstream DC load short-circuit fault rather than a fault on the DC bus. The contributions of this paper are:

\begin{itemize}
    \item A comprehensive analysis of the load-side fault mechanism for LLC converters and a corresponding fault detection framework to instantly restrict the fault current.
    \item A unified current-limiting architecture with a closed-loop current regulator that maintains the DC current at a designated level, allowing fault isolation and resilient operation without additional hardware.
    \item A coordinated ramp design that jointly restores the switching frequency and duty cycle after isolation, bounding the inrush current and ensuring a soft recharge.
\end{itemize}

\begin{figure}[ht]
\centering
\includegraphics[width=0.74\textwidth]{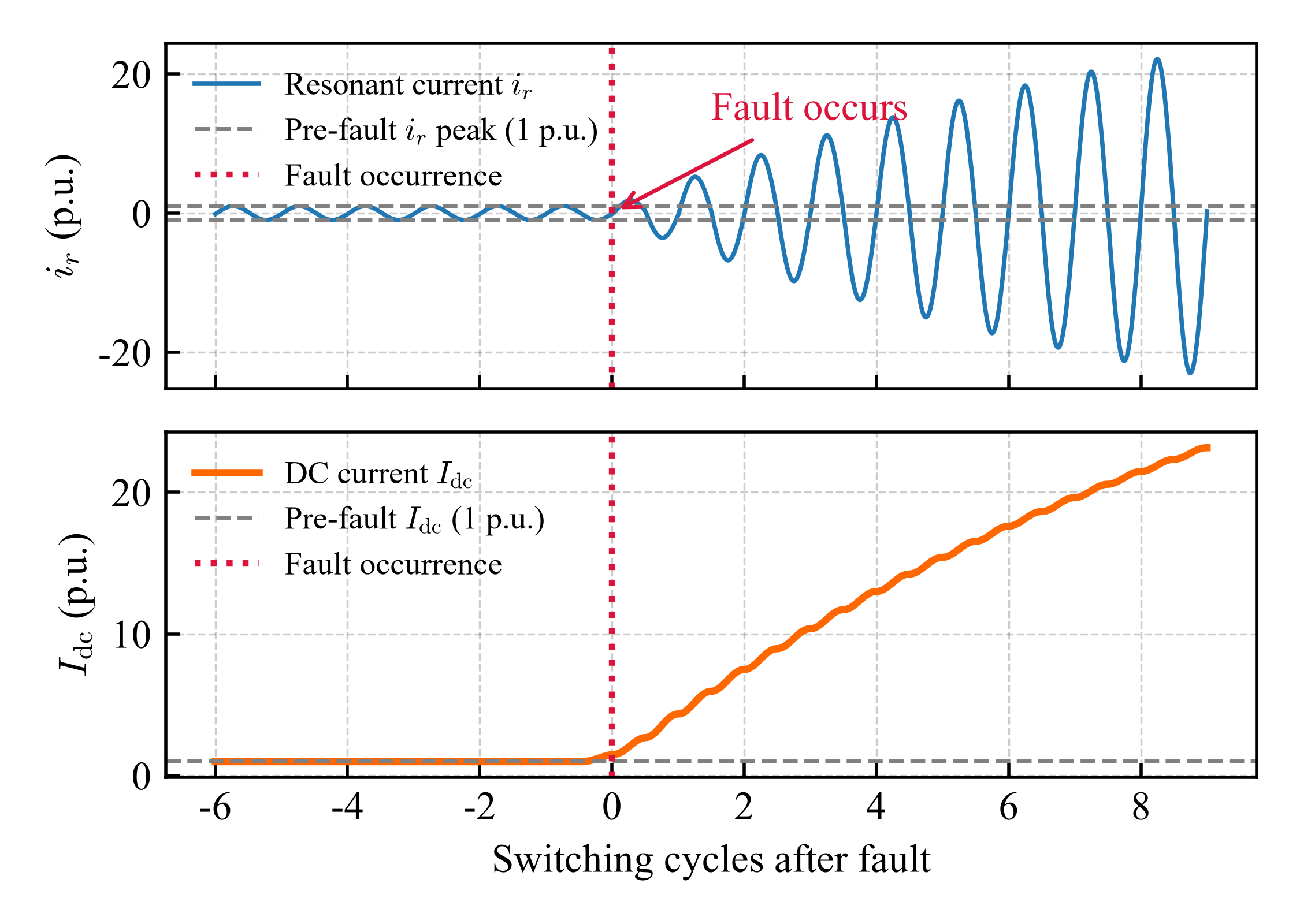}
\caption{The example fault currents of LLC converters without current limiting.}
\label{fig:fault_current}
\end{figure}

\begin{figure}[ht]
\centering
\includegraphics[width=\textwidth]{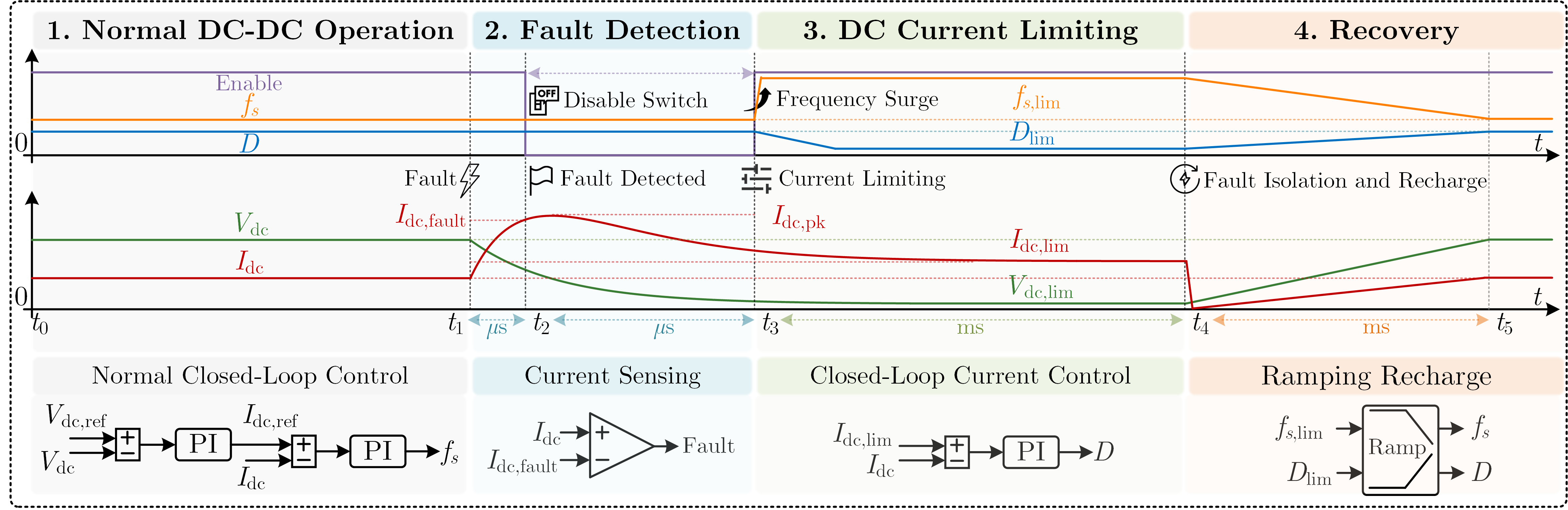}
\caption{Proposed DC fault-tolerant current-limiting and recovery strategy for LLC converters: (1) normal DC-DC operation, (2) fault detection, (3) DC current limiting, and (4) recovery, together with the corresponding waveforms of enable, $f_s$, $D$, $V_{\mathrm{dc}}$, and $I_{\mathrm{dc}}$, and the control block diagram used in each stage.}
\label{fig:strategy}
\end{figure}

\section{DC Fault Mechanism and Analysis}

An LLC DC-DC converter is studied as an example, as shown in Fig.~\ref{fig:system}(b). It comprises a full-bridge inverter $S_{1-4}$ with an input voltage $V_i$ that drives a resonant tank ($L_{r}$, $C_{r}$, $L_{m}$), a transformer $T$ of turns ratio $n_{1}\!:\!n_{2}$, and a full-bridge diode rectifier $D_{1-4}$ with a capacitor $C_{o}$. Suppose that a short-circuit occurs near loads, causing a small output resistance $R_{s}$ that reflects to the primary side as:
\begin{equation}
    R_{\mathrm{ac}} = \frac{8}{\pi^{2}}\left(\frac{n_{1}}{n_{2}}\right)^{2} R_{s}.
    \label{eq:Rac}
\end{equation}

During the fault, the second-order resonant tank behaves as:
\begin{equation}
    \frac{d^{2} i_{r}}{d t^{2}}
    + 2\alpha \frac{d i_{r}}{d t}
    + \omega_{r}^{2}\, i_{r}
    = \frac{1}{L_{r}}\frac{d v_{s}}{d t}.
    \label{eq:2ndorder}
\end{equation}

Define the resonant frequency, characteristic impedance, damped coefficient, damped frequency, and damping ratio as:
\begin{equation}
\begin{aligned}
    \omega_{r} &= 2\pi f_r=\sqrt{L_{r}C_{r}}^{-1},
    \quad
    Z_{0} = \sqrt{L_{r}/C_{r}},
    \quad \\
    \alpha &= \frac{R_{\mathrm{ac}}}{2 L_{r}},
    \quad
    \omega_{d} = \sqrt{\omega_{r}^{2} - \alpha^{2}},
    \quad
    \zeta = \frac{\alpha}{\omega_{r}}.
\end{aligned}
\end{equation}

With the switching frequency $f_{s} = f_{r}(1+\delta)$, the steady-state peak resonant and DC currents can be expressed as:
\begin{equation}
    \hat{i}_{r,\infty} \approx \frac{4V_{i}}{\pi\sqrt{R_{\mathrm{ac}}^{2} + 4\delta^{2}Z_{0}^{2}}}, \quad \hat{I}_{\mathrm{dc},\infty} = \frac{2}{\pi} \frac{n_{1}}{n_{2}}\,\hat{i}_{r,\infty}.
    \label{eq:ihat_ss}
\end{equation}

The transient overshoot for the underdamped system is:
\begin{equation}
    M_{p} = \exp\!\left(-\frac{\pi \zeta}{\sqrt{1-\zeta^{2}}}\right)
    \approx e^{-\pi \zeta}.
    \label{eq:Mp}
\end{equation}

The peak transient currents are approached at $t=\pi/\omega_d$:
\begin{equation}
    \hat{i}_{r,\mathrm{pk}} \approx \hat{i}_{r,\infty}\left(1 + e^{-\pi \zeta}\right), \quad
    \hat{I}_{\mathrm{dc},\mathrm{pk}} \approx \hat{I}_{\mathrm{dc},\infty}\left(1 + e^{-\pi \zeta}\right).
    \label{eq:ipeak}
\end{equation}

Other types of DC-DC converters can be analyzed in similar ways. Generally, a small $R_{\mathrm{s}}$ will lead to large peak currents in a very short time, causing severe system damage, as shown in Fig.~\ref{fig:fault_current}. Thus, a current limiting strategy is strongly required.

\section{Current Limiting and Recovery Strategy}

The proposed current limiting strategy, as illustrated in Fig.~\ref{fig:strategy}, is implemented entirely within a local Digital Signal Processor (DSP) and requires no communication with downstream protection devices, making it suitable for fast autonomous fault ride-through. Generally, it consists of 4 main stages: normal operation, fault detection, current limiting, and recovery, with each described below for LLC converters.

\subsection[Stage 1: Normal DC-DC Operation]{Stage 1: Normal DC-DC Operation ($t_0 \le t < t_1$)}

Under normal conditions, DC-DC operation of SSTs usually adopts multifarious closed-loop control strategies according to the specific topology. Typically, the DC-DC stage regulates the DC voltage $V_{\mathrm{dc}}$ to its reference $V_{\mathrm{dc,ref}}$ using a cascaded voltage-current closed-loop control.

During this stage, the duty cycle $D$ of the switched-node voltage $v_s$ is maintained at its nominal value (i.e., typically $D=1$) while the frequency modulation regulates the voltage.

\subsection[Stage 2: Fault Detection]{Stage 2: Fault Detection ($t_1 \le t < t_3$)}

When a short-circuit load fault occurs at $t_1$, resulting in low equivalent impedance $R_s$ of the DC bus, the DC current $I_{\mathrm{dc}}$ rises sharply as analyzed, which can be utilized as an intuitive and accurate indicator of fault occurrence. Therefore, fault detection can be continuously implemented at hardware level by analog comparators or at software level by current sensing, then it is detected by condition:
\begin{equation}
\mathcal{F}=\left( I_{\mathrm{dc}} > I_{\mathrm{dc,fault}} \right) \cap \left( \mathrm{d}I_{\mathrm{dc}}/\mathrm{d}t > \zeta_{\mathrm{fault}} \right)
\end{equation}
where $I_{\mathrm{dc,fault}}$ is the fault threshold of the DC current and $\zeta_{\mathrm{fault}}$ is the threshold of its increasing rate. Considering $I_{\mathrm{dc}}$ is sensed in the original control, the software-level comparison can be implemented without additional hardware.
For example, for each data from current sensors, the analog-digital conversion including the sensor response takes \SI{5}{\micro\second} while the transmission through optical fibers takes another \SI{5}{\micro\second}. Therefore, it takes \SI{10}{\micro\second} for the DSP to receive the instantaneous current information, which is acceptable for SSTs with low frequencies (e.g., \SI{100}{kHz}).

When the fault is flagged at $t_2$, the DSP immediately disables switches to prevent $I_{\mathrm{dc}}$ and $i_{r}$ from being too high until $I_{\mathrm{dc}}$ is discharged to a system-friendly level before the current-limiting controller takes authority at $t_3$.

\subsection[Stage 3: DC Current Limiting]{Stage 3: DC Current Limiting ($t_3 \le t < t_4$)}

After a duration of switch disabling, the converter needs to maintain a certain amount of current $I_{\mathrm{dc,lim}}$ under the low impedance condition to melt fuses and isolate the fault branch. Therefore, at $t_3$, the controller transitions to the proposed current limiting strategy that contains two coordinated parts.

\subsubsection{Frequency Surge:}

For the LLC converter, the voltage conversion ratio is:
\begin{equation}
M=\frac{n_1 V_{\mathrm{dc}}}{n_2 V_i}= \frac{k\,\sin\!\left(\dfrac{D\pi}{2}\right)}{\sqrt{\left[(k+1) - \dfrac{1}{f_n^2}\right]^2 + \left[\dfrac{k Z_0}{R_{\mathrm{ac}}}\left(f_n - \dfrac{1}{f_n}\right)\right]^2}}
\label{eq:voltage_conversion_ratio}
\end{equation}

where $k = L_m/L_r$, $f_n = f_s/f_r$, and the transmitted power is $P=V_{\mathrm{dc}}^{2}/R_{s}$ while the required power is $P_{\mathrm{lim}}=I_{\mathrm{dc,lim}}^{2}R_s$.

Notably, $M(f_n,D)$ decreases with $f_n$. Therefore, the current limiting strategy commands that $f_s$ is stepped up rapidly to a limiting value $f_{s,\mathrm{lim}}$ (e.g., $f_{s,\mathrm{lim}}=2f_r$), immediately reducing the instantaneous power capability and providing a fast first line of current reduction. Moreover, the frequency surge reduces the switching period, making the duty cycle control introduced below more feasible in real controllers.

\subsubsection{Closed-Loop Current Control via Duty Cycle}

As illustrated by \eqref{eq:voltage_conversion_ratio}, $M(f_n,D)$ can change significantly with $D$, where a small $D$ leads to a very low $M$ that is suitable for the current-limiting case. The variation of $D$ can be achieved by adding an internal phase shift between $S_1$ and $S_4$ in LLC converters or equivalent actions in other types of DC-DC converters, and provides a feasible solution to regulating the current during fault with closed-loop control. The proposed current loop is structurally analogous to traditional current control but referenced to $I_{\mathrm{dc,lim}}$, and ramps the duty cycle from its default value toward $D_{\mathrm{lim}}$:
\begin{equation}
D = K_{p,\mathrm{lim}}\big(I_{\mathrm{dc,lim}} - I_{\mathrm{dc}}\big) + K_{i,\mathrm{lim}}\!\int \big(I_{\mathrm{dc,lim}} - I_{\mathrm{dc}}\big)\,dt .
\label{eq:limit_loop}
\end{equation}

The combined action of the frequency surge and the current loop regulates $I_{\mathrm{dc}}$ to a constant $I_{\mathrm{dc,lim}}$ during the fault, while $V_{\mathrm{dc}}$ is allowed to collapse toward a reduced $V_{\mathrm{dc,lim}}$. This stage typically persists for milliseconds, allowing protection devices to isolate the fault branch.

\subsection[Stage 4: Recovery]{Stage 4: Recovery ($t \ge t_4$)}

Once the fault is cleared at $t_4$, $I_{\mathrm{dc}}$ drops to a very small value due to the restored resistance and $V_{\mathrm{dc}}$ begins to rise. The controller detects the fault isolation and switches to recharge the DC bus and restore healthy operation. Re-energizing the bus directly at nominal operation would risk an inrush current, causing potential device damage. The proposed strategy instead applies rate-limited ramps to $f_s$ and $D$ for soft charging:
\begin{equation}
f_s(t) = f_{s,\mathrm{lim}} - k_f \,(t - t_4), \quad
D(t) = D_{\mathrm{lim}} + k_D \,(t-t_4)
\label{eq:ramp}
\end{equation}
where $k_f$ and $k_D$ are the ramp rates, and $f_s$ and $D$ are clamped between the limiting and nominal values.
This produces a controlled monotonic recharge of $V_{\mathrm{dc}}$ and $I_{\mathrm{dc}}$ back to nominal values in milliseconds, after which control authority is returned to the normal closed-loop operation.

\section{Experimental Verification}

\begin{figure}[tbp]
\centering
\includegraphics[width=0.76\textwidth]{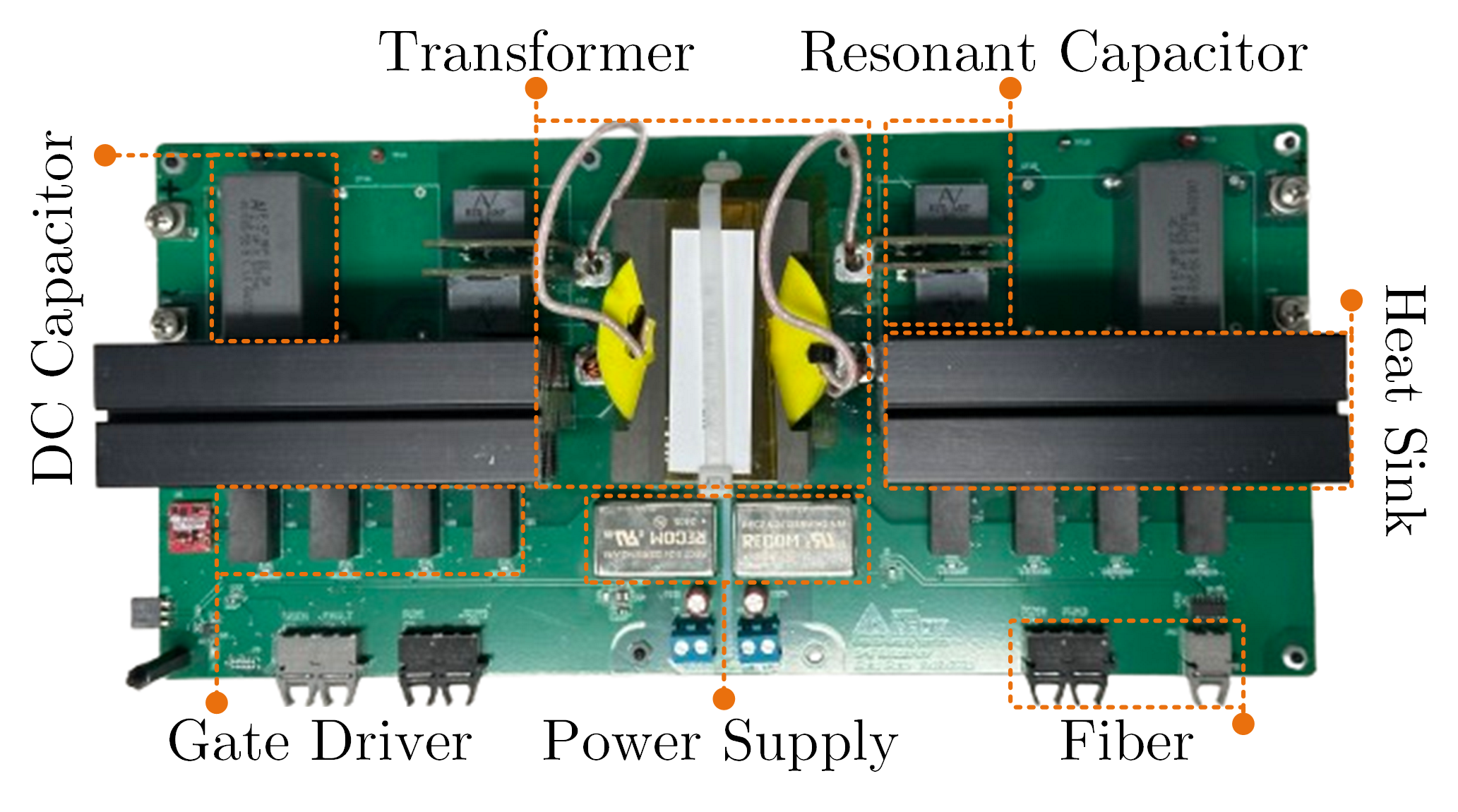}

\caption{The experimental prototype of LLC converters.}
\label{fig:prototype}
\end{figure}

\begin{table}[tbp]
\centering
\caption{Main parameters of the experimental setup.}
\label{tab:parameters}
\small
\renewcommand{\arraystretch}{1.2}
\begin{tabular*}{\textwidth}{@{\extracolsep{\fill}}cccccc@{}}
\toprule
\rowcolor{gray!15}
Parameter & Value & Parameter & Value & Parameter & Value \\
\midrule
$V_i$ & \SI{400}{V} & $V_{\mathrm{dc,ref}}$ & \SI{400}{V} & $f_r$ & \SI{154}{kHz} \\
$R_L$ & \SI{80}{\ohm} & $R_s$ & \SI{2}{\ohm} & $f_{s,\mathrm{lim}}$ & \SI{300}{kHz} \\
$L_r$ & \SI{7}{\micro\henry} & $L_m$ & \SI{70}{\micro\henry} & $n_1:n_2$ & $1:1$ \\
$C_o$ & \SI{3}{\micro\farad} & $C_r$ & \SI{0.15}{\micro\farad} & $I_{\mathrm{dc,lim}}$ & \SI{15}{A} \\
\bottomrule
\end{tabular*}
\end{table}

Experiments based on an LLC converter prototype, as shown in Fig.~\ref{fig:prototype}, are conducted to verify the feasibility of the proposed strategy. The main parameters of the converter are illustrated in Table \ref{tab:parameters} and the switches are IMZ120R030M1H SiC MOSFET. Specifically, the LLC converter works with $V_i=V_{\mathrm{dc}}=$\SI{400}{V} and $R_L=$\SI{80}{\ohm} under normal conditions (i.e., the output power is \SI{2}{kW}). A small resistance $R_s=$\SI{2}{\ohm} is placed to simulate the DC fault for illustrative purposes.

\begin{figure}[ht]
\centering
\includegraphics[width=\textwidth]{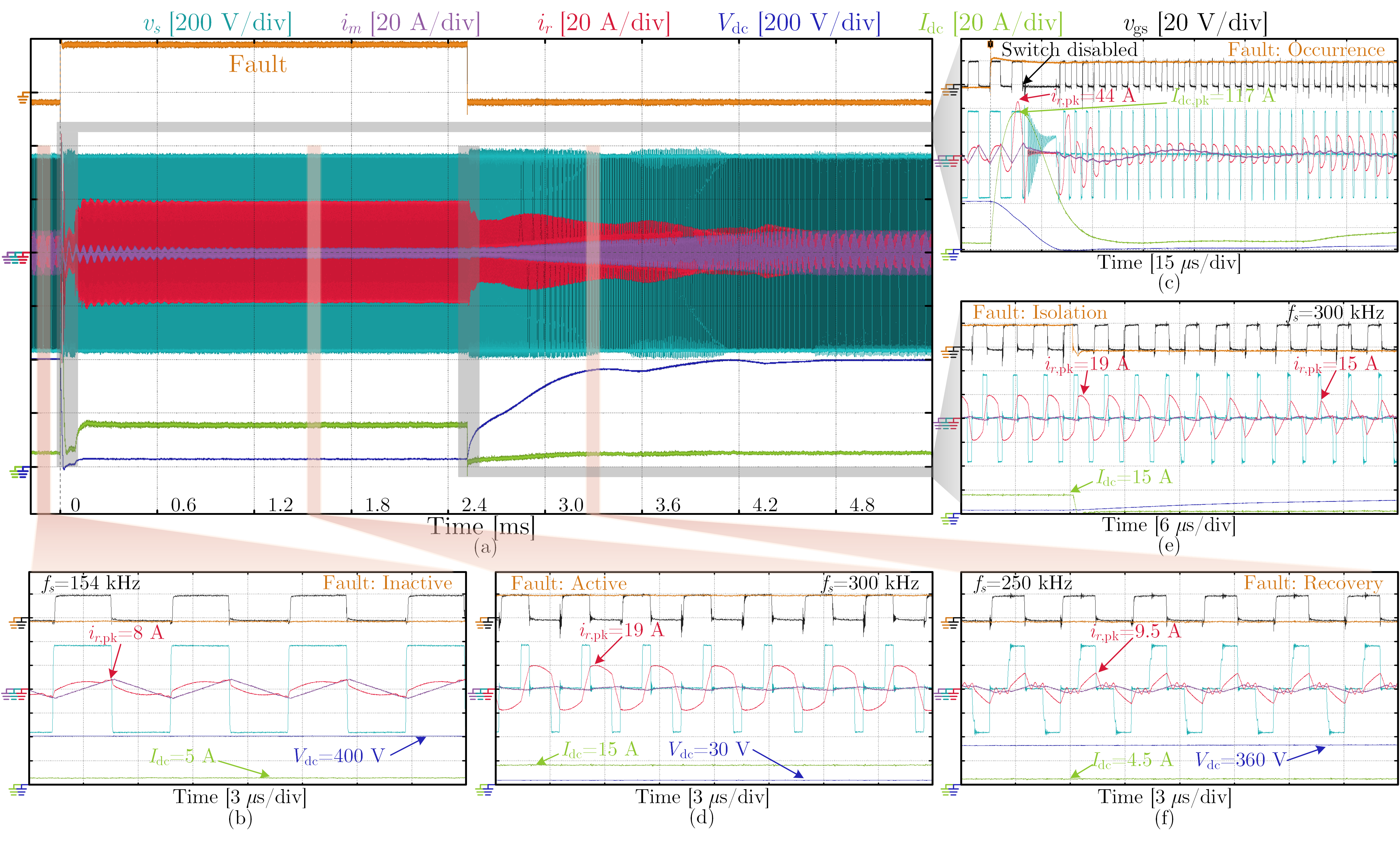}
\caption{Experimental verification of the proposed strategy on an LLC converter. (a) Overall performance. (b) Normal operation without fault. (c) Transient period of fault occurrence. (d) Current-limiting stage. (e) Transient period of fault isolation. (f) One period in recovery after fault.}
\label{fig:experiment}
\end{figure}

The experimental results are shown in Fig.~\ref{fig:experiment}. Specifically, Fig.~\ref{fig:experiment}(a) shows the overall performance of the converter during the DC fault transition, where the converter shows an exceptional capability to handle the fault for safety and provide the required current for fault isolation. Fig.~\ref{fig:experiment}(b) illustrates the normal waveforms where $f_s=f_r=$ \SI{154}{kHz}, $V_{\mathrm{dc}}=$\SI{400}{V}, and $I_{\mathrm{dc}}=$\SI{5}{A}. The DC fault occurs (i.e., $R_s$ is connected) in Fig.~\ref{fig:experiment}(c), forcing $V_{\mathrm{dc}}$ to decrease and $I_{\mathrm{dc}}$ and $i_{r}$ to increase rapidly and reach peaks of \SI{117}{A} and \SI{44}{A} in \SI{10}{\micro\second}, respectively. Note that when $I_{\mathrm{dc}}$ exceeds $I_{\mathrm{dc,fault}}=$\SI{20}{A}, the controller disables the switches for the next switching cycle, preventing the currents from becoming fatally high. After $C_o$ is fully discharged, the controller increases $f_s$ to \SI{300}{kHz} and switches to the closed-loop current control for current limiting. As shown in Fig.~\ref{fig:experiment}(d), $I_{\mathrm{dc}}$ is stably controlled to $I_{\mathrm{dc,lim}}=$\SI{15}{A} by adjusting the duty cycle $D$ of $v_s$ and $V_{\mathrm{dc}}$ stays at \SI{30}{V}. The whole stage lasts for several milliseconds, offering enough time for fuses to melt. In Fig.~\ref{fig:experiment}(e), the fault is isolated, and
$f_s$ and $D$ remain the same initially and change with the designed ramp rates for the subsequent recovery, as shown in Fig.~\ref{fig:experiment}(f).

The experiments verify that the LLC converter can provide sufficient currents under the proposed current-limiting strategy during the fault and recover under ramped switching frequency and duty cycle without inrush currents in the resonant tank, thus offering a feasible solution to riding through faults.

\section{Conclusion}

This paper presented a fault-tolerant current-limiting and recovery strategy for the DC-DC stage of solid-state transformers serving DC data center buses, especially voltage-source types. Specifically, the proposed strategy detects the fault by current sensing and firmware-level comparison, and limits the DC fault current to a designated value for milliseconds by coordinating a switching-frequency surge with a dedicated duty-cycle-based closed-loop current controller, so that fuses or circuit breakers can isolate the faulted load branch and keep the SST enabled. A subsequent coordinated rate-limited recovery stage then restores nominal operation without excessive inrush currents during the recharge. The proposed strategy is implemented at the control-loop level and entirely within the local controller, and it requires no additional hardware cost and no communication with downstream protection devices, making it well suited to the speed and reliability requirements of resilient data center power architectures. Experiments on an LLC converter have verified that the proposed strategy can restrict the massive fault current spikes, limit the current for protection devices, and restore the bus without inrush currents.

\FloatBarrier
\begingroup
\small
\bibliographystyle{IEEEtran}
\bibliography{references}

@ARTICLE{5634051,
  author={Huang, Alex and others},
  journal={Proceedings of the IEEE}, 
  title={The Future Renewable Electric Energy Delivery and Management ({FREEDM}) System: The Energy Internet}, 
  year={2011},
  volume={99},
  number={1},
  pages={133-148},
}

@ARTICLE{6648465,
  author={Deng, Junjun and others},
  journal={IEEE Transactions on Vehicular Technology}, 
  title={Design Methodology of {LLC} Resonant Converters for Electric Vehicle Battery Chargers}, 
  year={2014},
  volume={63},
  number={4},
  pages={1581-1592},
}

@ARTICLE{9343854,
  author={Weng, Haoyuan and others},
  journal={IEEE J. Emerg. Sel. Top. Power Electron.}, 
  title={A {DC} Solid-State Transformer With {DC} Fault Ride-Through Capability}, 
  year={2022},
  volume={10},
  number={4},
  pages={3617-3630},
}

@ARTICLE{4118335,
  author={Xie, Xiaogao and others},
  journal={IEEE Trans. on Power Electron.}, 
  title={Analysis and Optimization of {LLC} Resonant Converter With a Novel Over-Current Protection Circuit}, 
  year={2007},
  volume={22},
  number={2},
  pages={435-443},
}

@ARTICLE{1331495,
  author={Meyer, C. and others},
  journal={IEEE Trans. on Power Electron.}, 
  title={Solid-state circuit breakers and current limiters for medium-voltage systems having distributed power systems}, 
  year={2004},
  volume={19},
  number={5},
  pages={1333-1340},
}

@INPROCEEDINGS{1179276,
  author={Yang, B. and others},
  booktitle={Eighteenth Annual IEEE Applied Power Electronics Conference and Exposition, 2003 APEC}, 
  title={Over current protection methods for {LLC} resonant converter}, 
  year={2003},
  volume={2},
  number={},
  pages={605-609},
}

@ARTICLE{9893537,
  author={Wang, Haoyu and others},
  journal={IEEE Trans. on Ind. Electron.}, 
  title={{ZVS} Soft Switching Operation Region Analysis of Modular Multi Active Bridge Converter Under Single Phase Shift Control}, 
  year={2023},
  volume={70},
  number={7},
  pages={6865-6875},
}

@Article{app12010015,
AUTHOR = {Perea-Mena, Bayron and others},
TITLE = {Circuit Breakers in Low- and Medium-Voltage {DC} Microgrids for Protection against Short-Circuit Electrical Faults: Evolution and Future Challenges},
JOURNAL = {Applied Sciences},
VOLUME = {12},
YEAR = {2022},
NUMBER = {1},
ARTICLE-NUMBER = {15},
PAGES = {15},
ISSN = {2076-3417},
}

@ARTICLE{9526632,
  author={Kheirollahi, Reza and others},
  journal={IEEE J. Emerg. Sel. Top. Power Electron.}, 
  title={Coordination of Ultrafast Solid-State Circuit Breakers in Radial {DC} Microgrids}, 
  year={2022},
  volume={10},
  number={4},
  pages={4690-4702},
}

@INPROCEEDINGS{6953876,
  author={Liu, Shuo and others},
  booktitle={2014 IEEE Energy Conversion Congress and Exposition (ECCE)}, 
  title={Short-circuit current control strategy for full-bridge {LLC} converter}, 
  year={2014},
  volume={},
  number={},
  pages={3496-3503},
}

@ARTICLE{7482715,
  author={Li, Rui and others},
  journal={IEEE Trans. Energy Convers.}, 
  title={Active Control of {DC} Fault Currents in {DC} Solid-State Transformers During Ride-Through Operation of Multi-Terminal {HVDC} Systems}, 
  year={2016},
  volume={31},
  number={4},
  pages={1336-1346},
}

@ARTICLE{10093060,
  author={Mou, Di and others},
  journal={IEEE Transactions on Power Electronics}, 
  title={Reactive Power Minimization for Modular Multi-Active-Bridge Converter With Whole Operating Range}, 
  year={2023},
  volume={38},
  number={7},
  pages={8011-8015},
}

@ARTICLE{10242261,
  author={Wang, Haoyu and others},
  journal={IEEE Transactions on Industrial Electronics}, 
  title={Universal Phase-Shift Modulation Scheme and Efficiency Optimization for Modular Multiactive Bridge Converter}, 
  year={2024},
  volume={71},
  number={7},
  pages={7312-7321},
}

@ARTICLE{10132875,
  author={Wang, Haoyu and Ji, Shiqi and Mou, Di and others},
  journal={IEEE Journal of Emerging and Selected Topics in Power Electronics}, 
  title={Switching Characterization and Power Loss Optimization for Modular Multiactive Bridge Converter Under Common Phase Shift Control}, 
  year={2023},
  volume={11},
  number={4},
  pages={3924-3936},
}

@ARTICLE{10194423,
  author={Mou, Di and others},
  journal={IEEE Transactions on Industrial Electronics}, 
  title={High-Efficiency Time-Division Multiplexing Modulation Technology for Modular Multiactive Bridge Converters}, 
  year={2024},
  volume={71},
  number={6},
  pages={5745-5754},
}

@ARTICLE{10319663,
  author={Wang, Haoyu and others},
  journal={IEEE Journal of Emerging and Selected Topics in Industrial Electronics}, 
  title={High-Frequency-Link-Based Reactive-Power Optimal Control for Modular Multi-Active Bridge Converter}, 
  year={2024},
  volume={5},
  number={3},
  pages={1333-1337},
}

@INPROCEEDINGS{10362302,
  author={Wang, Haoyu and Mou, Di and Ji, Shiqi and others},
  booktitle={2023 IEEE Energy Conversion Congress and Exposition (ECCE)}, 
  title={Comparison and Improvement of ZVS Operation Under Different Modulation Strategies for Modular Multi Active Bridge Converters}, 
  year={2023},
  volume={},
  number={},
  pages={2556-2563},
}

@INPROCEEDINGS{9628583,
  author={Xiang, Minjiang and others},
  booktitle={2021 IEEE 1st International Power Electronics and Application Symposium (PEAS)}, 
  title={Design and Analysis of a 75 kVA High-Frequency-Link Based Three-Port Power Electronic Transformer}, 
  year={2021},
  volume={},
  number={},
  pages={1-7},
}

@ARTICLE{11358392,
  author={Zheng, Jialin and others},
  journal={IEEE Transactions on Industrial Electronics}, 
  title={Physics-Embedded Neural ODEs for Sim-to-Real Edge Digital Twins of Hybrid Power Electronics Systems}, 
  year={2026},
  volume={73},
  number={6},
  pages={8616-8627},
}

@ARTICLE{11077776,
  author={Zheng, Jialin and Wang, Haoyu and Zeng, Yangbin and others},
  journal={IEEE Transactions on Industrial Electronics}, 
  title={Cognitive Digital Twins-Based Model Predictive Control for High-Frequency Power Converters}, 
  year={2025},
  volume={72},
  number={12},
  pages={13310-13321},
}

@INPROCEEDINGS{10583830,
  author={Wang, Haoyu and Ji, Shiqi and Mou, Di and Zeng, Yangbin},
  booktitle={2024 IEEE 7th International Electrical and Energy Conference (CIEEC)}, 
  title={Model-Free Approximate Fundamental Reactive Power Minimization for Modular Multi-Active Bridge Converter}, 
  year={2024},
  volume={},
  number={},
  pages={2068-2073},
}
\endgroup

\end{document}